\documentclass[a4paper, conference]{IEEEtran}
\usepackage{algorithm}
\usepackage{algpseudocode}

\usepackage{cite}
\usepackage{amsmath,amssymb,amsfonts}

\usepackage{graphicx}
\usepackage{textcomp}
\usepackage{xcolor}

\usepackage{amsthm}

\begin{document}

\title{Hybrid Deep Reinforcement Learning for Joint Resource Allocation in Multi-Active RIS-Aided Uplink Communications \\
} 

\author{\IEEEauthorblockN{ Mohamed M. H. Shalma}
\IEEEauthorblockA{\textit{Faculty of IET} \\
\textit{German university in Cairo}\\
mohamed.hamed@guc.edu.eg}
\and
\IEEEauthorblockN{ Engy Aly Maher}
\IEEEauthorblockA{\textit{Faculty of IET} \\
\textit{German university in Cairo}\\
engy.aly@guc.edu.eg}
\and
\IEEEauthorblockN{ Ahmed El-Mahdy}
\IEEEauthorblockA{\textit{Faculty of IET} \\
\textit{German university in Cairo}\\
ahmed.elmahdy@guc.edu.eg}
}

\maketitle

\begin{abstract}
Active Reconfigurable Intelligent Surfaces (RIS) are a promising technology for 6G wireless networks. This paper investigates a novel hybrid deep reinforcement learning (DRL) framework for resource allocation in a multi-user uplink system assisted by multiple active RISs. The objective is to maximize the minimum user rate by jointly optimizing user transmit powers, active RIS configurations, and base station (BS) beamforming. We derive a closed-form solution for optimal beamforming and employ DRL algorithms—Soft actor-critic (SAC), deep deterministic policy gradient (DDPG), and twin delayed DDPG (TD3)—to solve the high-dimensional, non-convex power and RIS optimization problem. Simulation results demonstrate that SAC achieves superior performance with high learning rate leading to faster convergence and lower computational cost compared to DDPG and TD3. Furthermore, the closed-form of optimally beamforming enhances the minimum rate effectively.
\end{abstract}

\begin{IEEEkeywords}
Reconfigurable Intelligent Surfaces (RIS), Deep Reinforcement Learning (DRL), Soft Actor-Critic (SAC), Deep Deterministic Policy Gradient (DDPG)
\end{IEEEkeywords}

\section{Introduction}
The higher need for better spectral efficiency, enhanced coverage, and energy-efficiency for 6G wireless systems has derived innovative research into smart radio environments. The reconfigurable intelligent surface (RIS) has emerged as a key technology capable of dynamically shaping electromagnetic waves through programmable meta-material elements. Unlike passive RIS which only reflects the signal with phase-shifting circuits, the recent evolution toward active RIS overcomes these constraints by integrating reflection-type amplifiers into the reflecting elements (RE)s enabling joint amplitude and phase adjustment of reflected signals. This amplification property significantly improves signal-to-noise ratio and mitigates the double-fading effect which makes it superior compared to passive ones \cite{9377648,9928771,9998527}. However, the intensive 6G requirements imposes higher challenges making research heading towards deploying multiple active RISs. The deployment of double active RIS is studied in some works using traditional optimization techniques \cite{9839181,10525765,10623414,10525785}. The coordination of multi-active RIS systems introduces unprecedented challenges: high-dimensional control variables, multiple dynamic noises, and more complicated interference suppression. Conventional optimization techniques (e.g., block coordinate descent, semi-definite relaxation) struggle with these complexities due to non-convexity, real-time processing requirements, and scalability limitations. Deep Reinforcement Learning (DRL) offers a promising alternative by leveraging neural networks to approximate high-dimensional policies through environment interactions. Its ability to learn optimal strategies in partially observable, stochastic wireless environments makes it uniquely suited for active RISs.
\section{Related Works}
The active RIS DRL-based optimization is investigated in several works \cite{10601864,10820336,10445710}. The authors showed the positive impact of optimizing active RISs using DDPG for maximizing the rate and energy efficiency. The DRL-based active RIS applications is also studied for different applications including cognitive systems \cite{10636057} and vehicular communication \cite{10992287}. However, all these works focus on the single RIS optimization. To the best of our knowledge, only \cite{10543479} considered optimizing multiple active RISs. The authors considered the resource allocation of a downlink system by optimizing the BS beamforming and active RISs elements to maximize the sum-rate using deep Q network (DQN) approach. In fact, the DQN handles discrete action spaces but cannot optimize continuous variables. In uplink scenarios, the users transmit power is an important resource which has to be optimized to mange interference. However, power optimization constructs a continuous action space which can not be optimized using DQN. Furthermore, the sum-rate maximization objective in \cite{10543479} can not guarantee fairness among users. In contrast to \cite{10543479}, we aim to maximize the minimum achievable rate of all users which is more complex but a more fair objective. This is done by optimizing the users transmit power, beamforming vectors at the BS, and multi-active RISs elements. Due to the more challenging objective and high dimensions of decision variables, we propose a hybrid DRL approach and use the SAC due its entropy regularization which leads to better exploration and escaping sub-optimal policies. Finally, we compare the proposed approach with DDPG used in \cite{10601864,10820336,10445710} and TD3 used in \cite{10636057}. The DDPG is an actor–critic algorithm designed for continuous action spaces. It uses a deterministic policy to map states directly to actions and employs separate neural networks for the actor (policy) and critic (value function). The algorithm also uses target networks and experience replay to stabilize training. Despite its effectiveness, DDPG is known to suffer from overestimation bias and sensitivity to hyper-parameters. The TD3 is an enhanced version of DDPG by using two Q-networks to reduce overestimation bias by taking the minimum of two critic estimates. summarize the differences between our work and some related works in TABLE I. 
\begin{table}
    \normalsize 
    \centering
\caption{Related works Comparison}
\label{tab:my_label}
    \begin{tabular}{|p{0.5cm}|p{0.8cm}|p{1.37cm}|p{1.1cm}|p{2.8cm}|c|c|c|c|c|}\hline
         Ref&  Power opt.&  Beamfor- ming opt.&  Multiple RISs& DRL algorithms\\\hline
         \cite{10601864} &  X&  $\checkmark$&  X& DDPG \\\hline
         \cite{10820336} &  X&  $\checkmark$&  X&DDPG \\\hline
         \cite{10445710} & X&  $\checkmark$  &  X& DDPG - PPO \\\hline
         \cite{10636057} & X&  $\checkmark$  &  X& DDPG - TD3 \\\hline
         \cite{10992287} & $\checkmark$ & $\checkmark$ &  X& DDPG \\\hline
         \cite{10543479} &  X& $\checkmark$ &  $\checkmark$& DQN\\ \hline
 Our & $\checkmark$& $\checkmark$& $\checkmark$&SAC - DDPG -TD3\\\hline
    \end{tabular}

\end{table}

The contributions of this paper are summarized as:
\begin{itemize}
    \item We investigate the performance of a multi-user, multi-active RIS system by jointly optimizing the transmit power, beamforming of the base station (BS) and active REs to maximize the minimum achievable rate.
    \item We provide a closed-form expression of the optimal beamforming at the BS.
    \item We compare the performance of the SAC with TD3 and DDPG proposed in related works as benchmarks.
\end{itemize}
\section{System Model and Problem Formulation}
We consider an uplink scenario as shown in Fig.1 where $L$ active RISs exist to assist the communications between $K$ users and the BS. Each RIS has a controller that communicates with the BS via a dedicated link to acknowledge the optimal REs that are calculated at the BS. Each user is equipped with a single antenna and the BS is equipped with $M$ antennas. The power of the $k$-th user is denoted by $p_k$ and its transmitted symbol is denoted by $s_k$. The BS  applies a beamforming vector $\boldsymbol{w}_k \in \mathbb{C}^{M \times 1}$ to decode the $k$-th user signal. We denote the beamforming matrix as $\boldsymbol{W}=[\boldsymbol{w}_1 ; \boldsymbol{w}_2 ; \cdots ; \boldsymbol{w}_K]$. Each active RIS is a planar array with a number of reflecting elements that can introduce both gains and phase shifts to the signal impinging on it. The number of REs for the $l$-th RIS is denoted by $N_l$  and $N_{total}=\sum^L_{l=1}{N_l}$ is the total number of active REs in the communication system. The direct channel between the $k$-th user and the BS is denoted by $\boldsymbol{d}_k \in \mathbb{C}^{M\times1}$. The channel between the $k$-th user and the $l$-th RIS is denoted by $\boldsymbol{h}_{k \rightarrow l} \in \mathbb{C}^{N_l \times 1}$. The channel between the $l$-th RIS and the BS is denoted by $\boldsymbol{G}_l \in \mathbb{C}^{M \times N_l}$. The noise vector of the $l$-th active RIS is denoted by $\boldsymbol{n}_l \in \mathbb{C}^{N_l \times 1}$ with variance $\sigma_l^2$ and the noise vector at the BS is denoted by $\boldsymbol{n}_0 \in \mathbb{C}^{M \times 1}$ with variance $\sigma_0^2$. $\boldsymbol{\Phi}_l=\text{diag}(\boldsymbol{\phi}_l)$ is the diagonal matrix of the active reflection coefficients of the $l$-th RIS. We define some sets as $\mathcal{K} \overset{\Delta}{=} \left\{ 1,2,\cdots,K\right\}$ , $\mathcal{L} \overset{\Delta}{=} \left\{ 1,2,\cdots,L\right\}$, $\mathcal{N} \overset{\Delta}{=} \left\{ 1,2,\cdots,N_{total}\right\}$.
\begin{figure}[!t]
    \centering
    \includegraphics[width=\linewidth]{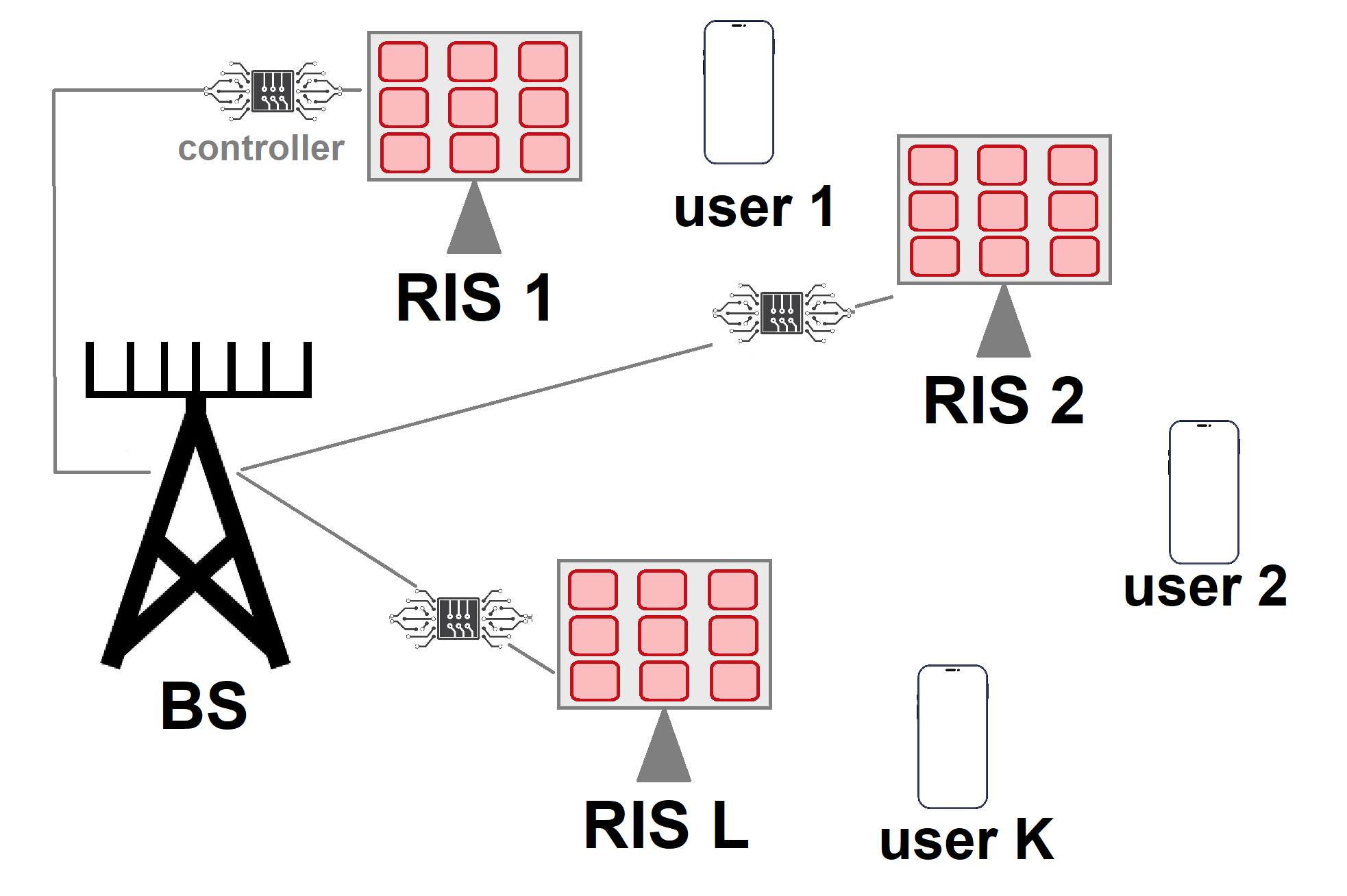}
    \caption{System Model}
    \label{fig:enter-label}
\end{figure}
We assume that the channel state information is known and constant during the time slot of transmission. The direct channels are estimated by turning off the RIS in the absorbing mode \cite{9722893}. The channels between the users and RISs and channels between RISs and the BS can be estimated using active sensors either using the cascaded or separate techniques \cite{9722893}. This arises from the fact that the channels $\boldsymbol{G}_l \ \forall l \in \mathcal{L}$ are generally quasi-static since the RISs and the BS are fixed and the active nature of active RISs \cite{9722893}. The channels are expressed using both large scale and small scale fading models as \vspace*{-0.3cm}
\begin{equation}\label{ch1}
    \boldsymbol{d}_{k}= \underbrace{\sqrt{\eta \ z_{d,k}^{-\zeta}}}_{\text{large scale fading}} ( \underbrace{  \alpha\boldsymbol{d}_{k}^{LoS} + \beta\boldsymbol{d}_{k}^{NLoS} }_{\text{small scale fading}} )
\end{equation}\vspace*{-0.5cm}
\begin{equation}\label{ch2}
    \boldsymbol{h}_{k \rightarrow l}= \underbrace{\sqrt{\eta \ z_{k \rightarrow l}^{-\zeta}}}_{\text{large scale fading}} ( \underbrace{  \alpha\boldsymbol{h}_{k \rightarrow l}^{LoS} + \beta\boldsymbol{h}_{k \rightarrow l}^{NLoS} }_{\text{small scale fading}} )
\end{equation}\vspace*{-0.5cm}
\begin{equation}\label{ch3}
   \boldsymbol{G}_l=\underbrace{\sqrt{\eta \ z_l^{-\zeta}}}_{\text{large scale fading}} ( \underbrace{\alpha\boldsymbol{G}_l^{LoS} + \beta\boldsymbol{G}_l^{NLoS}}_{\text{small scale fading}} )
\end{equation}
\begin{figure*}[!t]
\begin{equation} \label{sinr}
 \gamma_k=   \frac{ p_k \left|  \boldsymbol{w}_k^H \left( \boldsymbol{d}_k +\sum\limits^L_{l=1}{ \boldsymbol{G}_l \boldsymbol{\Phi}_l \boldsymbol{h}_{k \rightarrow l} } \right) \right|^2 }    {  \sum\limits^K_{i=1,i \neq k}{ p_i \left| \boldsymbol{w}_k^H \left( \boldsymbol{d}_i + \sum\limits^L_{l=1}{ \boldsymbol{G}_l \boldsymbol{\Phi}_l \boldsymbol{h}_{i \rightarrow l} }  \right) \right|^2 }  +  \sum\limits^L_{l=1}{\sigma_l^2 \| \boldsymbol{w}_k^H \boldsymbol{G}_l \boldsymbol{\Phi}_l \|^2}  + \sigma_0^2 \| \boldsymbol{w}_k \|^2}
\end{equation}
\hrulefill
\end{figure*}
where $\eta$ is the path-loss gain at reference distance of one meter, $z_{d,k}$ is the distance between the $k$-th user and the BS, $z_{k \rightarrow l}$ is the distance between the $k$-th user and the $l$-th RIS, $z_l$ is the distance between the $l$-th RIS and the BS, $\zeta$ is the path-loss exponent. The small scale fading follows a Ricain distribution with both line-of-sight with factor $\alpha=\sqrt{ {K_R}/{(1+K_R)} }$ and non-line-of-sight component with factor $\beta=\sqrt{ {1}/{(1+K_R)} }$ as demonstrated in (\ref{ch1})-(\ref{ch3}) where $K_R$ is the Rician factor.
The received signal at the BS is expressed as
\begin{multline}\label{y_signal}
      y_k=\boldsymbol{w}_k^H \underbrace{\sum^K_{k=1}{\boldsymbol{d}_k \sqrt{p_k} s_k}}_{\text{direct signals}}   +  \boldsymbol{w}_k^H\underbrace{\sum^K_{k=1}{\sum^L_{l=1}{ \boldsymbol{G}_l \boldsymbol{\Phi}_l \boldsymbol{h}_{k \rightarrow l} \sqrt{p_k} s_k}}}_{\text{reflected signals}} \\
      + \boldsymbol{w}_k^H\underbrace{\sum^L_{l=1}{  \boldsymbol{G}_l \boldsymbol{\Phi}_l \boldsymbol{n}_l  } }_{\text{dynamic noise of RISs}}+\boldsymbol{w}_k^H{\boldsymbol{n}_0}
\end{multline}
The SINR of the $k$-th user is given by (\ref{sinr}). The normalized rate of the $k$-th user is given by $R_k=\log_2(1+\gamma_k)$. Our objective is to maximize the minimum rate which is formulated as
\begin{equation*}
\mathcal{P}_1: \quad\quad   \max_{\boldsymbol{p},\boldsymbol{W},\boldsymbol{\Phi}_1,\boldsymbol{\Phi}_2,\cdots,\boldsymbol{\Phi}_L} \quad  \min \quad R_k
\end{equation*}
subject to
\begin{gather}
C_1 \quad\quad\quad\quad\quad\quad\quad        {\| \boldsymbol{w}_k \|^2} \leq 1 \ \ \ \forall k \in \mathcal{K} \quad  \label{c_Ew} \\
C_2  \quad\quad\quad\quad\quad\quad\quad  p_k \leq p_{max} \ \ \ \ \forall k \in \mathcal{K} \quad  \label{c_p}\\
C_3      \quad \left| \phi_l[n] \right| \leq \phi_l^{max} \ \ \ \ \forall l \in \mathcal{L} \ , \ n \in \left\{1,2,\cdots, N_l \right\}  \label{c_gain}\\
C_4  \quad   \angle {\phi_l[n]} \ \in [0,2\pi)   \ \forall l \in \mathcal{L} \ ,  \ n \in \left\{1,2,\cdots, N_l \right\} \label{c_phase}
\end{gather}
Constraint $C_1$ restricts the power consumed by the BS during the beamforming process, Constraint $C_2$ restricts the transmit power of each user below the maximum limit $p_{max}$ .  Constraint $C_3$ restricts the gain of each active RE to its maximum $\phi_l^{max}$ and constraint $C_4$ restricts the phase shift of each active RE between $0$ and $2\pi$. The difficulty in solving $\mathcal{P}_1$ lies in:
\begin{itemize}
    \item \textbf{Non-Convexity:} $\mathcal{P}_1$ is non-convex due to the non-convexity of the objective function and the coupling of the optimization variables. Furthermore, the max-min rate objective is non-smooth making gradient estimation difficult.
    \item \textbf{High-dimensional action space:} the number of decision variables of $\mathcal{P}_1$ is $O(M \times K+K+N_{total})$ which is equivalent to an action space of size $\mathcal{P}_1$ from $O(2 \times M \times K+K+2 \times N_{total})$ as deep networks take complex variables in real and imaginary form. At large $M,K,N_{total}$, solving $\mathcal{P}_1$ using usual optimization techniques only or even machine learning techniques only can yield extraordinary complexity, hardness of convergence to optimal solution due to the extreme large variables dimensions, extreme hardware resources and being stuck in local optimal policies.
    \item \textbf{Dynamic environment:} Wireless channels are time-varying due to user mobility, obstacles, and interference. Optimal RIS configurations must adapt to these changes and should be quick.
\end{itemize}
Due to the non-convexity, high dimensions and real-complex nature of  $\mathcal{P}_1$,  we propose a hybrid approach where $\mathcal{P}_1$ is decomposed into tractable two problems where one of them is solved using standard optimization techniques while the other one is solved using DRL-based approach.
\section{Proposed Optimization}
In this section, we propose an efficient hybrid approach using optimization theory and DRL. We first decompose $\mathcal{P}_1$ into two subproblems i) beamforming vectors optimization and ii) joint power and active REs optimization. This hybrid approach not only derive closed form expression of the optimal beamforming as shown later, but also reduces the action space of $\mathcal{P}_1$ from $O(2 \times M \times K+K+2 \times N_{total})$ to $O(K+2 \times N_{total})$ via decoupling the beamforming vectors and power, active REs.
\subsection{Beamforming Optimization in closed-form}
The beamforming optimization subproblem is given by
\begin{equation*}
\mathcal{P}_2: \quad\quad   \max_{\boldsymbol{W}} \quad  \min \left\{ R_k\right\}
\end{equation*}
subject to
\begin{center}
    $C_1$
\end{center}
we first start with a closed-form optimal expression for the beamforming. For brevity, let us define the equivalent channel of user $k$ from the direct and RISs links as
\begin{equation}
    \boldsymbol{ h}_k^{\text{eqv}}=\boldsymbol{d}_k +\sum\limits^L_{l=1}{ \boldsymbol{G}_l \boldsymbol{\Phi}_l \boldsymbol{h}_{k \rightarrow l} } 
\end{equation}
it is noted that the beamforming vectors are independent of each other as $\boldsymbol{w}_k$ affects the $k$-th user rate only and does not affect other users rates. Therefore, $  \max \left\{\log_2(1+\gamma_k)\right\} = \max \left\{\gamma_k\right\}$. The SINR can be re-written as a generalized Rayleigh quotient with respect to $\boldsymbol{w}_k$ as
\begin{equation}\label{sinr_rayleigh}
    \gamma_k=\frac{ \boldsymbol{w}_k^H \left(p_k  \boldsymbol{ h}_k^{\text{eqv}} (\boldsymbol{ h}_k^{\text{eqv}})^H \right)\boldsymbol{w}_k} { \boldsymbol{w}_k^H  \boldsymbol{\Lambda}_k \boldsymbol{w}_k }
\end{equation}
where $\boldsymbol{\Lambda}_k$ is given by
\begin{equation}
\boldsymbol{\Lambda}_k=\sum\limits_{i=1, i\neq k} p_i  \boldsymbol{ h}_i^{\text{eqv}} (\boldsymbol{ h}_i^{\text{eqv}})^H  +\ \sum\limits^L_{l=1}{\sigma_l^2  \boldsymbol{G}_l \boldsymbol{\Phi}_l \boldsymbol{\Phi}_l^H \boldsymbol{G}_l^H}  + \sigma_0^2 \boldsymbol{I}_M 
\end{equation}
the optimal solution to maximize (\ref{sinr_rayleigh}) is given by
\begin{equation}
    \boldsymbol{w}_k^{opt}=\sqrt{p_k}\boldsymbol{\Lambda}_k^{-1} \boldsymbol{h}_k^{\text{eqv}}
\end{equation}
which aims at minimizing the denominator of (\ref{sinr_rayleigh}) (via $\boldsymbol{\Lambda}_k^{-1}$) while maximizing the numerator (via $\boldsymbol{h}_k^{\text{eqv}}$). to abide by constraint $C_1$, the optimal solution is scaled by its magnitude to satisfy the unity constraint $C_1$
\begin{equation}
    \boldsymbol{w}_k^{opt}=\frac{\sqrt{p_k}\boldsymbol{\Lambda}_k^{-1} \boldsymbol{h}_k^{\text{eqv}}}{\| \sqrt{p_k}\boldsymbol{\Lambda}_k^{-1} \boldsymbol{h}_k^{\text{eqv}} \|}
\end{equation}

\subsection{Power and Active REs Optimization via DRL}
In this subsection, we use DRL to find the optimal active REs. The joint power and active REs optimization subproblem is formulated as
\begin{equation*}
\mathcal{P}_3: \quad\quad   \max_{\boldsymbol{p},\boldsymbol{\Phi}_1,\boldsymbol{\Phi}_2,\cdots,\boldsymbol{\Phi}_L} \quad  \min \left\{ R_k \right\}
\end{equation*}
subject to
\begin{center}
    $C_2 - C_4$
\end{center}
The DRL- based solution is mainly composed of observations, action, DRL algorithm, and reward function. These components are further illustrated as:\\
\textbf{Observations}: The observation is a continuous vector. Deep networks take only real numbers as input thus, the channel observations are passed as real and imaginary components which is defined as 
\begin{multline}
\texttt{obsv}=\Big[\operatorname{Re}\left\{ \boldsymbol{w}_k \boldsymbol{d}_k \right\}|_{\forall k \in \mathcal{K}};\operatorname{Im}\left\{ \boldsymbol{w}_k \boldsymbol{d}_k \right\}|_{\forall k \in \mathcal{K}}; \\
\operatorname{Re}\left\{ \boldsymbol{w}_k \boldsymbol{G}_l \text{diag}(\boldsymbol{h}_{k \rightarrow l}) \right\}|_{\forall k \in \mathcal{K}, l \in \mathcal{L}};\\
\operatorname{Im}\left\{\boldsymbol{w}_k \boldsymbol{G}_l \text{diag}(\boldsymbol{h}_{k \rightarrow l}) \right\}|_{\forall k \in \mathcal{K}, l \in \mathcal{L}} \Big]
\end{multline}
\textbf{Action Space}: The action space is a continuous vector containing both the  active REs and power. The action is vector defined as 
\begin{multline}
\hspace*{-0.3cm}\texttt{act}=\Big[\underbrace{p_1;p_2;\cdots;p_K}_{\text{power}} \ ; \  \underbrace{\operatorname{Re}\left\{ \boldsymbol{\phi_1}\right\};\operatorname{Re}\left\{ \boldsymbol{\phi_2}\right\};\cdots;\operatorname{Re}\left\{ \boldsymbol{\phi_L}\right\}}_{\text{real part of active REs}};\\
\underbrace{\operatorname{Im}\left\{ \boldsymbol{\phi_1}\right\};\operatorname{Im}\left\{ \boldsymbol{\phi_2}\right\};\cdots;\operatorname{Im}\left\{ \boldsymbol{\phi_L}\right\}}_{\text{imaginary part of active REs}}\Big]
\end{multline}
The action space has dimension of $K+2 \times N_{total}$. Constraint $C_2$ can be handled by limiting the power values in the action space vector as deep networks elements have upper and lower bounds. However, the active REs constraint $C_3$ has to be reformulated as upper/lower constraints for the DRL network to handle them. Accordingly, we transform $C_3$ as
\begin{equation}\label{c_changed}
     \quad \left| \phi_l[n] \right| \leq \phi_l^{max} \iff \operatorname{Re}\{\phi_l[n]\}^2+\operatorname{Im}\{\phi_l[n]\}^2 \leq (\phi_l^{max})^2
\end{equation}
from (\ref{c_changed}) we observe the upper bound of $\operatorname{Re}\{\phi_l[n]\}$ is $\phi_l^{max}$ and its lower bound is $-\phi_l^{max}$ when $\operatorname{Im}\{\phi_l[n]\}$ is $0$. The same applies for the imaginary part $\operatorname{Im}\{\phi_l[n]\}$ which has upper bound of $\phi_l^{max}$ and lower bound of $-\phi_l^{max}$ as well. To make sure $ \left| \phi_l[n] \right| \leq \phi_l^{max}$ strictly, a normalization layer is added in the DRL network.\\
\textbf{Agent}: We use the SAC agent. Unlike DDPG and TD3, SAC main point of strength is that it maximizes the expected return and the entropy of the policy as
\begin{equation}
J(\pi) = \sum_{t} \mathbb{E}_{(s_t, a_t) \sim \rho_\pi} \big[ \texttt{reward} - \alpha^{SAC}\underbrace{\mathbb{E}_{a_t \sim \pi}[\log \pi(a_t | s_t)}_{\text{entropy}}]\big]
\end{equation}
where $\alpha^{SAC}$ is the parameter controlling the entropy trade-off, $t$ is time slot. The actor and soft-Q (critic) are updated respectively as
\begin{equation}
J_\pi(\phi) = \mathbb{E}_{s_t \sim \mathcal{D}} \left[ 
    \mathbb{E}_{a_t \sim \pi_\phi} \left[ 
        \alpha \log \pi_\phi(a_t|s_t) - Q_\theta(s_t, a_t) 
    \right] 
\right]
\end{equation}
\begin{equation}
    Q(s_t, a_t) = r(s_t, a_t) + \gamma  \mathbb{E}_{s_{t+1} \sim p} \left[ 
    V(s_{t+1}) 
\right]
\end{equation}
This entropy incorporation enables superior exploration in high-dimensional action spaces and avoids local optima. \\
\textbf{Reward}: The reward is given to the agent each time it selects an action which relates to the objective function thus, the reward function is defined as $\texttt{reward} = \min{R_k}$. The overall optimization is described in Algorithm 1.

\begin{algorithm}[t]
\caption{Hybrid Resource Optimization}
\begin{algorithmic}[1]
\State \textbf{Initialize:} $p_{\max}$, channel matrices $\{\mathbf{d}_k\}, \{\mathbf{h}_{k \rightarrow l}\}, \{\mathbf{G}_l\}$, SAC agent parameters.
        \For{each user $k = 1$ to $K$}
            \State Compute equivalent channel $\boldsymbol{h}_k^{\text{e}}$ and $\boldsymbol{\Lambda}_k$
            \State Compute $\mathbf{w}_k^{\text{opt}}$ as $\mathbf{w}_k^{\text{opt}} = \frac{\sqrt{p_k} \boldsymbol{\Lambda}_k^{-1} \mathbf{h}_k^{\text{eqv}}}{\| \sqrt{p_k} \boldsymbol{\Lambda}_k^{-1} \mathbf{h}_k^{\text{eqv}} \|}$
        \EndFor
\For{each episode}
    \State Initialize environment, observe initial state $s_0$
    \For{each time step $t$}
        \State SAC agent selects action $a_t = [p_1, \dots, p_K, \Re\{\boldsymbol{\phi}_l\}, \Im\{\boldsymbol{\phi}_l\}]$
        \State Update RIS coefficients $\boldsymbol{\Phi}_l$ and user powers $p_k$ based on $a_t$
        \State Compute user rates $R_k = \log(1 + \gamma_k)$
        \State Compute reward: $r_t = \min_k R_k$
        \State Observe next state $s_{t+1}$, store transition $(s_t, a_t, r_t, s_{t+1})$
        \State Update SAC agent using replay buffer and policy/value loss
        \If{convergence criteria met}
            \State \textbf{Break}
        \EndIf
    \EndFor
\EndFor
\State \textbf{Output:} Optimized $\{\mathbf{p}_k\}|_{\forall k \in \mathcal{K}}$, $\{\boldsymbol{\Phi}_l\}|_{\forall l \in \mathcal{L}}$, $\{\mathbf{w}_k^{\text{opt}}\}|_{\forall k \in \mathcal{K}}$
\end{algorithmic}
\end{algorithm}

\section{Simulation Results}
In this section, we provide the numerical results of our simulation. The simulation parameters are provided in TABLE II unless otherwise stated. The minimum achievable rate vs the number of BS antennas is shown Fig. 2 for the proposed closed-form expression compared with random beamforming. The optimal beamforming shows a monotonic increase in the minimum rate with the number of antennas. As the number of antennas increase, the gap between optimal and random beamforming increases the minimum rate from 1 at $M=4$ to around 5 at $M=16$.\\

The minimum rate is plotted in Fig. 3 vs the number of episodes for learning rate of $10^{-2}$. We assume four RISs each with 10 elements and four users. The SAC can approach a highly optimal solution easily under 100 episodes. The DDPG and TD3 on the other hand stuck in sub-optimal policies where the TD3 achieves only 25$\%$ of the rate achieved by SAC meanwhile DDPG fails. As mentioned before, SAC relies on entropy regularization which increases its exploration compared to DDPG and TD3. Although high exploration leads to some minimal oscillations as shown in Fig. 3, it can easily escape sub-optimal policies at high learning rates enabling fast convergence with less computational complexity.\\
\begin{figure}
    \centering
    \includegraphics[width=\linewidth]{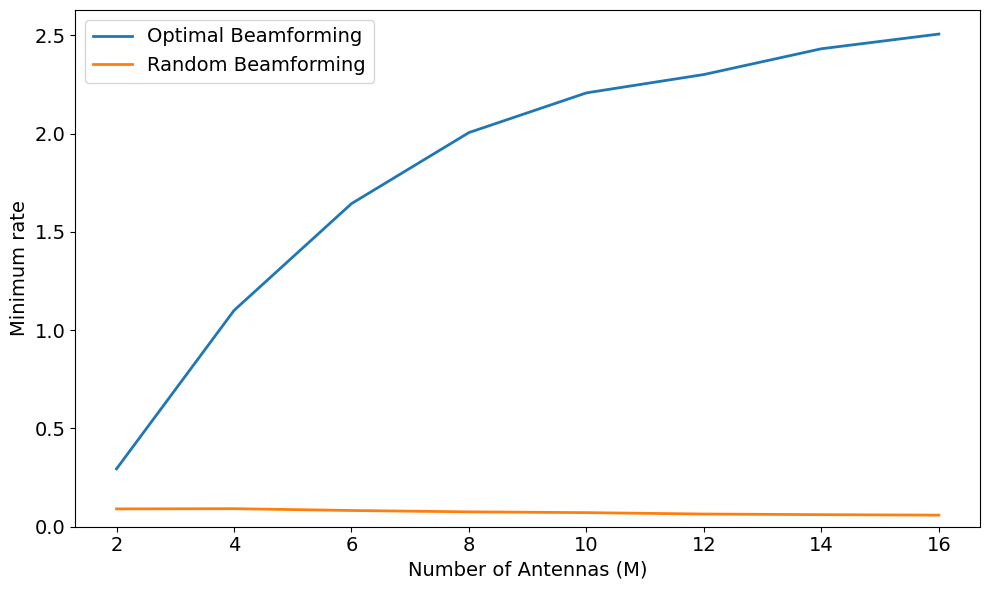}
    \caption{Minimum SINR vs number of BS antennas}
    \label{fig:enter-label}
\end{figure}
\begin{figure}
    \centering
    \includegraphics[width=\linewidth]{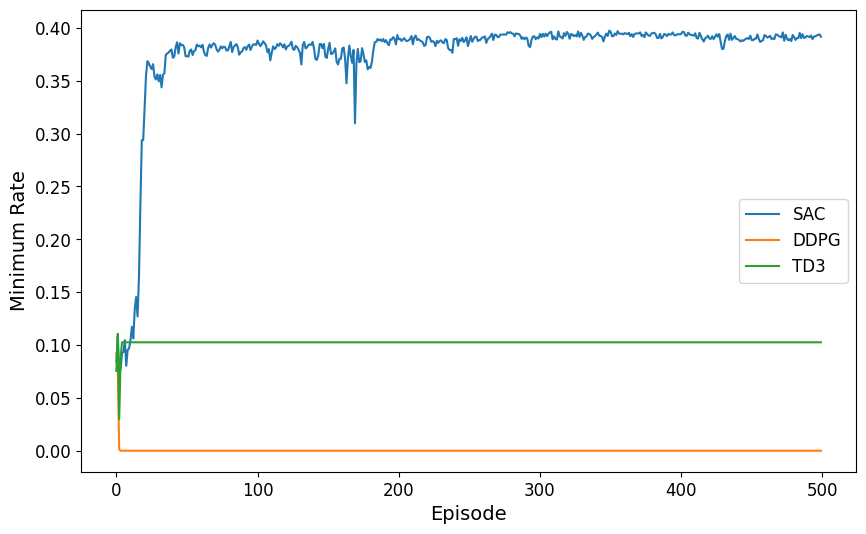}
    \caption{Minimum rate vs number of episodes for SAC, DDPG, and TD3 with learning rate=$10^{-2}$}
    \label{fig:enter-label}
\end{figure}

\begin{table}
\normalsize
\centering
\caption{Simulation Parameters}
\label{tab:my_label}
\begin{tabular}{|c|c|}
\hline
\quad\quad\quad \textbf{Parameter} \quad\quad\quad & \quad\quad\quad \textbf{Value} \quad\quad\quad \\\hline
Learning Rate & $10^{-2}$ \\\hline
Buffer Size & $10^5$ \\\hline
Batch Size & 128 \\\hline
Soft Update Parameter & $10^{-3}$ \\\hline
Discount Factor & 0.99 \\\hline
No. of steps per episode & 50 \\\hline
No. hidden layers & 2 \\\hline
No. of neurons & 128 \\\hline
 No. of active RISs&4\\\hline
 No. of users&4\\\hline
max transmit power & 23 dBm \\\hline
max RIS gain & 3 dB \\\hline
BS location & (0,0,5) \\\hline
RISs distance from BS & 12m \\\hline
users distance from BS & 6m \\\hline
\end{tabular}
\end{table}

The minimum rate vs the number of episodes is plotted in Fig. 4 for learning rate of $10^{-3}$. With lower learning rate, the performance of DDPG and TD3 initially enhances and reaches a mid-optimal policy due to their reliance on small learning rates. However, when increasing the number of episodes, the DDPG maintains the sub-optimal solution and TD3 fails. On the other hand, SAC maintains the same optimal solution obtained with learning rate of $10^{-2}$ (which is 0.4 bits/sec/Hz) and still outperform both DDPG and TD3 specifically, when the number of episodes exceeds 150.\\

To fully explore different DRL algorithms sensitivity to the learning rate, we plot the minimum rate vs the number of episodes in Fig. 5 for learning rate of $10^{-4}$. Below 1000 episodes, both DDPG and TD3 highly outperform SAC as SAC is still in the exploration phase. However, both DDPG and TD3 can not maintain full stability and severe degradation of DDPG occurs at episode 450. The TD3 shows more swinging behavior and more degradation than DDPG. On the other hand, SAC converges to the optimal policy after 1800 episodes but shows a stable behavior without severe drops.\\

In brief, results from Fig. 3, 4, and 5 shows that DDPG and TD3 can outperform SAC initially if the learning rate is small. On the other hand, SAC performs better when the learning rate is fast and maintains better overall stability and resilience to being stuck in sub-optimal policies due to its entropy regularization and high exploration.\\

To explore SAC convergence under different situations, we plot the minimum rate vs the number of episodes in Fig. 6 for three scenarios: i) four users and four RISs each with a number of elements equal to 10 ($N_{total}=40$), ii) six users and four RISs each with a number of elements equal to 20 ($N_{total}=80$), and iii) eight users and four RISs each with a number of elements equal to 100 ($N_{total}=400$). The learning rate is set to $0.5\times 10^{-2}$. Case (i) approaches the optimal policy after 70 episodes and experiences minimal swinging and after 150 episodes nearly stabilizes. Case (ii) approaches the optimal policy after 170 episodes and also experiences minimal swinging and nearly stabilizes after 210 episodes. For case (iii) with the worst complexity and action space of size $2 \times 400 + 8 =808$ optimization variables, the minimum rate highly swings till 400 episodes an exhibits small swinging after that. After 720 episodes, it nearly stabilizes to its optimal policy.

\begin{figure}
    \centering
    \includegraphics[width=\linewidth]{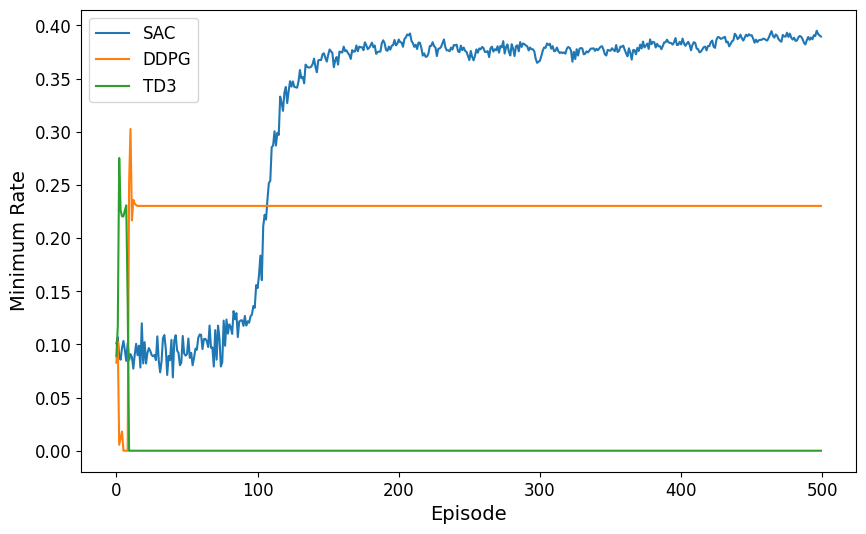}
    \caption{Minimum rate vs number of episodes for SAC, DDPG, and TD3 with learning rate=$10^{-3}$}
    \label{fig:enter-label}
\end{figure}
\begin{figure}
    \centering
    \includegraphics[width=\linewidth]{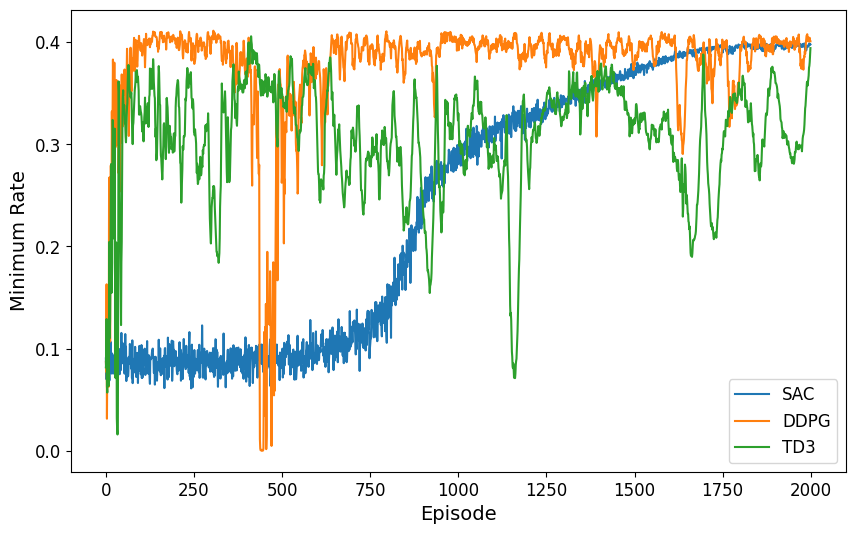}
    \caption{Minimum rate vs number of episodes for SAC, DDPG, and TD3 with learning rate=$10^{-4}$}
    \label{fig:enter-label}
\end{figure}

\begin{figure}
    \centering
    \includegraphics[width=\linewidth]{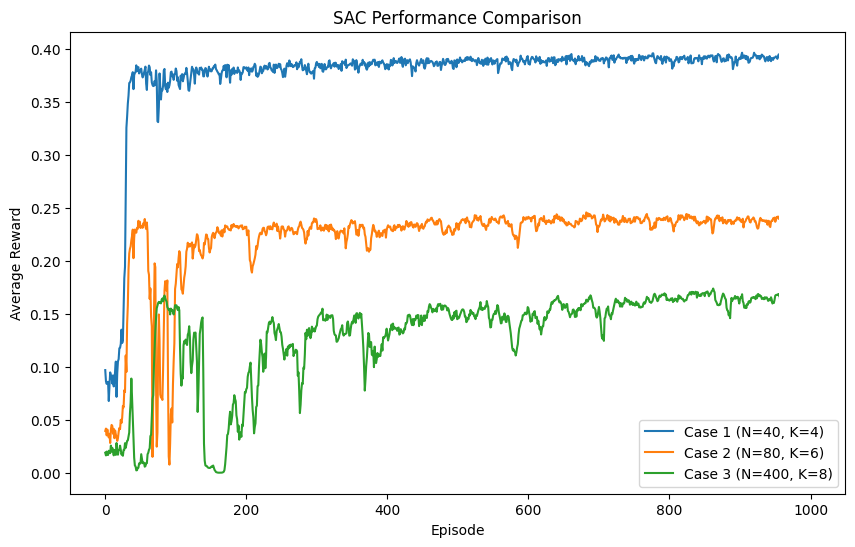}
    \caption{Minimum rate vs number of episodes using SAC for different $N,K$}
    \label{fig:enter-label}
\end{figure}
\section{Conclusion}
In this paper, we investigated the performance of multi-active RIS uplink system using hybrid DRL-approach. Results show that SAC converges with fast learning rate compared to DDPG and TD3 and easily escapes sub-optimal policies and that the proposed closed-form optimal beamforming significantly increases the minimum achievable rate.

\bibliographystyle{ieeetr}
\bibliography{Zref}

\end{document}